\documentclass[aps,prl,twocolumn,showpacs,preprintnumbers,amsmath,amssymb,floatfix]{revtex4-1}
\usepackage{graphicx}
\usepackage{amsfonts}
\usepackage{CJK}
\usepackage{bm}
\usepackage{color}
\usepackage{xcolor}
\usepackage{array}
\usepackage[T1]{fontenc}
\usepackage[colorlinks = true,linkcolor = blue,urlcolor = blue,citecolor = blue,anchorcolor = blue]{hyperref}

\begin{document}
\bibliographystyle{apsrev4-1}

\title{\textcolor{black}{Optimally Fast Qubit Reset}}
\author{Yue Liu}\email{yueliu@xmu.edu.cn}\thanks{These authors contributed equally to this work.}
\author{Chenlong Huang}\thanks{These authors contributed equally to this work.}
\author{Xingyu Zhang}
\author{Dahai He}\email{dhe@xmu.edu.cn}

\affiliation{Department of Physics and Jiujiang Research Institute, Xiamen University, Xiamen 361005, Fujian, China}

\date{\today}
\begin{abstract}
  \textcolor{black}{In practice, qubit reset must be operated in an extremely short time, which incurs a thermodynamic cost within multiple orders of magnitude above the Landauer bound. We present a general framework to determine the minimal thermodynamic cost and the optimal protocol for arbitrary resetting speeds. Our study reveals the divergent behavior of minimal entropy production in the short-time limit depends on the convergence and divergence of the jump operators. For the convergent class, an inherent trade-off exists between the minimal required time and the set error probability, which hinders the Moore's law continuing in such cases. Moreover, we find the optimal protocol exhibits the similarity in the fast-driving regime for different times. To demonstrate our findings, we empoly fermionic and bosonic baths as examples. Our results suggest that the super-Ohmic bosonic heat bath is a suitable choice for qubit reset.}

\end{abstract}

\maketitle
    \textit{Introduction.}---Analyzing the energy consumption of quantum computers is essential for evaluating their performance constraints. Qubit reset is one of the DiVincenzo criteria for quantum computation~\cite{Di2000}. Landauer principle states that the energy cost to reset a qubit is bounded below by~\cite{Landauer1961, Bennett1982, Georgescu2021}
    \begin{equation}\label{Landauer}
        Q\geq \beta^{-1}\ln{2},
    \end{equation}
    where $\beta$ denotes the inverse temperature $\beta=1/(k_{\text{B}}T)$ of the environment. This fundamental principle sets a limit on the thermodynamic cost of information processing independent of the physical system or hardware~\cite{Parrondo2015}. The Landauer principle has been experimentally verified in various systems, such as colloidal particles in double-well potential~\cite{Antoine2012, Jun2014, Gavrilov2016}, trapped ultracold ion~\cite{Yan2018}, nanomagnets memory bit~\cite{Hong2016, Martini2016, Gaudenzi2018}, underdamped micromechanical oscillators~\cite{Dago2021, Dago2022, Dago2023}, and quantum dots~\cite{Scandi2022}.

    \textcolor{black}{The qubit reset is an essential step in quantum process and biochemical systems, such as the quantum metrology~\cite{Chu2022}, quantum sensing~\cite{Niroula2024} and ribosomes decoding computations~\cite{kempes2017}, which requires high speed and high accuracy. Unfortunately, the thermodynamic cost of qubit reset for both artificial and natural systems ~\cite{Wolpert2024} are many orders of magnitude above the Landauer bound which becomes a major challenge to the indefinite continuation of the Moore's law~\cite{Waldrop2016}.} Therefore, recent theoretical studies have focused on the finite-time memory erasure. Several finite-time Landauer bounds have been derived from thermodynamic bounds such as speed limits~\cite{Zhen2021, Tan20221, Zhen2022, Lee2022, Tan2023}. In addition, it has been shown that the lower bound of the additional cost grows linearly with the speed of reset~\cite{Proesmans2020L, Proesmans2020E, Zhen2021, Tan20221, Boyd2022, Konopik2023}, which has been observed in various phenomena of finite-time thermodynamics~\cite{Ma2020, Yuan2022}. On the other hand, much effort has been dedicated to the study of the minimal thermodynamic cost and the optimal protocol in the slow-driving regime~\cite{Ma2022, Scandi2022}. \textcolor{black}{Despite the remarkable advances, qubit reset in the fast-driving regime, which is the most crucial for practical applications, has become a missing piece of the puzzle in theoretical research. Experimentally, the efficiency of qubit reset has been significantly improved in recent years~\cite{Magnard2018,Zhou2021,Johnson2022,Reiner2024}, such as a fidelity exceeding 99\% within a duration of 300 ns~\cite{Alibaba2024}. To guide the experimental and practical design to continue the Moore's law, the deeper theoretical understanding in such a short time is urgently desired.}

    In this Letter, in order to investigate the fast qubit reset, we propose a general framework for analyzing the minimal thermodynamic cost and the optimal protocol across different timescales. \textcolor{black}{We show the divergent behavior of the minimal entropy production in the fast-driving regime depends on the convergence and divergence of the jump operators.} For the convergent class, we prove the existence of the minimal required time for a given error probability, which significantly limits the speed of reset in such class of environment. \textcolor{black}{Moreover, we demonstrate the similarities in the optimal protocols for different timescales in the short-time limit.} Our results, supported by various examples, provide valuable insights into qubit reset, such as quantum effects of environment.

\begin{figure}[ht]
      \centering
      \includegraphics[width=1\linewidth]{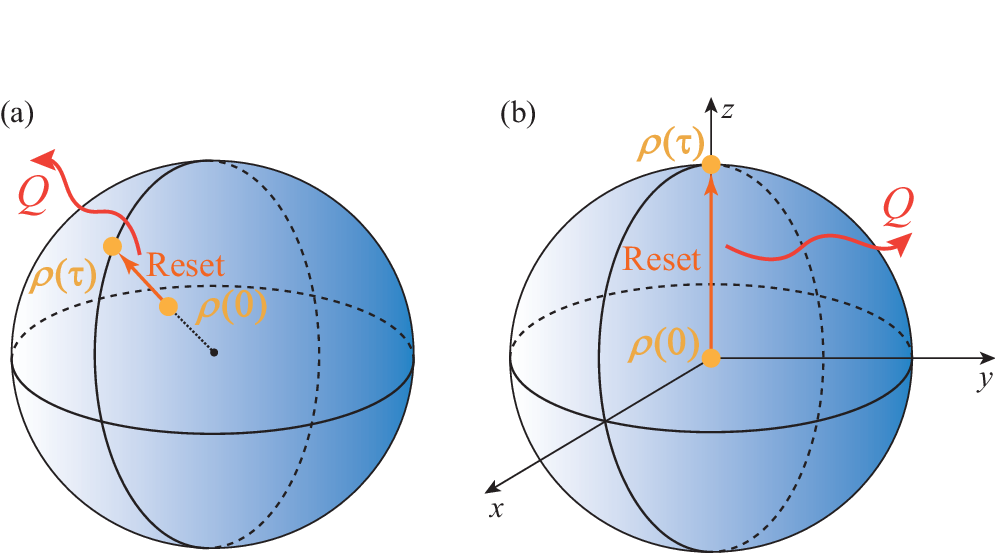}
      \caption{\label{Fig1} \textcolor{black}{Illustration of the qubit reset in the Bloch sphere. (a) From an arbitrary mixed state $\rho(0)$, the qubit is reset to the pure state $\rho(\tau)$ by the optimal protocol. (b) From the maximally mixed state, the qubit is reset to the ground state of the $z$-direction Pauli matrix by the optimal protocol, which is employed to demonstrate the memory erasure.}}
      \end{figure}

    \textit{Setup.}---\textcolor{black}{As illustrated in Fig.~\ref{Fig1}(a), the qubit reset involves erasing information encoded in a mixed state located within the Bloch sphere. The target final state is a pure state located on the surface of the Bloch sphere, in which the von Neumann entropy is zero. We consider the process realized by using a controllable qubit with Hamiltonian
    \begin{equation}\label{Hamiltonian}
      H_{t}=\frac{\lambda(t)}{2}\boldsymbol{\sigma}\cdot\boldsymbol{n}(t),
    \end{equation} 
    where $\lambda$, $\boldsymbol{\sigma}$ and $\boldsymbol{n}$ present the driving strength, the Pauli matrix and unit vector in the direction of driving, respectively.} For brevity, the Planck constant and the Boltzmann constant are set to unity $\hbar=k_{\text{B}}=1$. \textcolor{black}{The initial state can be an arbitrarily mixed state $\rho(0)$ represented by an inferior point of the Bloch sphere.} The qubit is then reset in a finite time $\tau$ by manipulating the Hamiltonian $H_{t}$. We focus on the Markov regime, where the qubit is weakly coupled to an infinite heat bath, described by the Lindblad master equation
    \begin{equation}\label{Lindblad}
      \dot{\rho}=-i[H_{t},\rho]+\sum_{k=0}^{1}L_{k}\rho L_{k}^{\dag}-\frac{1}{2}\{L_{k}^{\dag}L_{k},\rho\}.
    \end{equation}
    The jump operators $L_{k}$, which are determined by heat bath, satisfy the detailed balance condition $L_{1}=e^{-\beta\lambda/2}L_{0}^{\dag}$. \textcolor{black}{The final state $\rho(\tau)$ is an almost pure state located near the surface of the Bloch sphere, with an error probability caused by the finite-time effect.} The heat production can be calculated by $Q=-\int_{0}^{\tau}\text{Tr}(\dot{\rho}H_{t})dt$. According to the second law of thermodynamics, the heat production is bounded by $Q\geq-T\Delta S$, where $\Delta S:=\text{Tr}[\rho(\tau)\ln{\rho(\tau)}]-\text{Tr}[\rho(0)\ln{\rho(0)}]$ denotes the change in von Neumann entropy.
    
    Previous studies have demonstrated that quantum coherence can induce additional dissipation~\cite{Browne2014, Miller2020, Tan20221}, often termed ``quantum friction''~\cite{Plastina2014}. To minimize heat production, the optimal protocol should avoid generating quantum coherence. \textcolor{black}{As one can see in Fig~\ref{Fig1}(a), the driving direction $\boldsymbol{n}$ is maintained to the energy eigenstate of initial state to keep the coherence vanish and the reset process is along the radius. Therefore, the density matrix remains the diagonal form $\rho=p_{0}|0\rangle\langle0|+p_{1}|1\rangle\langle1|$, where $|0\rangle$ and $|1\rangle$ represent the ground and the excited state corresponding to the eigenstate of initial state. The maximally mixed state $\rho(0)=(1/2)|0\rangle\langle0|+(1/2)|1\rangle\langle1|$ located at the center of the Bloch sphere is the most difficult to reset, which is sufficient to understand the heat dissipation of the qubit reset~\cite{Tan2023}. As depicted in Fig~\ref{Fig1}(b), we employ the maximally mixed state as the initial state and fix the driving direction $\boldsymbol{n}$ along the $z$-axis to demonstrate the memory erasure process.}

    We consider the transition rate from excited state to ground state as $\langle1|L_{0}L_{0}^{\dag}|1\rangle=\mu R(\beta\lambda)$, where $R(\beta\lambda)$ represents rescaled transition rate and $\mu$ denotes the overall transition rate that is a constant in our study. The transition rate from the ground state to the excited state is $\langle0|L_{1}L_{1}^{\dag}|0\rangle=\mu R(\beta\lambda)e^{-\beta\lambda}$ because of the detailed balance condition. Finally, at $t=\tau$, the qubit evolves to an almost pure ground state $\rho(\tau)=(1-\varepsilon)|0\rangle\langle0|+\varepsilon|1\rangle\langle1|$ with the error probability $\varepsilon$.

    \textit{The minimal thermodynamic cost.}---We present the minimal energy cost and the corresponding optimal protocol by investigating the heat functional subjected to the boundary conditions $p_{1}(0)=1/2$ and $p_{1}(\tau)=\varepsilon$. By considering the master equation Eq.~\eqref{Lindblad}, we obtain the functional with respect to $p$
    \begin{equation}\label{heat_functional2}
        Q= \beta^{-1}\int_{\varepsilon}^{\frac{1}{2}}\tilde{\lambda}\left(p,\frac{{\rm d} \tilde{t}}{{\rm d} p}\right){\rm d} p,
    \end{equation}
    with $\tilde{\lambda} = \beta\lambda$, $\tilde{t}=t/\tau$ and $p(\tilde{t})=p_{1}(\tau\tilde{t})$. The optimal $\tilde{\lambda}(p)$ that minimizes heat production can be determined by solving the Euler-Lagrange equation with respect to Eq.~\eqref{heat_functional2}
    \begin{equation}\label{lp}
        CpR(\tilde{\lambda})(1-e^{-\zeta})^{2}-\frac{R'(\tilde{\lambda})}{R(\tilde{\lambda})}(1-e^{-\zeta})- e^{-\zeta}=0,
    \end{equation}
    where the constant $C$ can be calculated by boundary condition and $\zeta=\tilde{\lambda}-\ln{[(1-p)/p]}$~\cite{supp}. Solving this equation yields the inverse function $\tilde{t}(p)$ of the optimal $p(\tilde{t})$
    \begin{equation}\label{tp}
        \tilde{t}(p)=\frac{1}{\mu\tau}\displaystyle\int_{p}^{\frac{1}{2}}\dfrac{{\rm d}u}{uR[\tilde{\lambda}(u)](1 - e^{-\zeta})}.
    \end{equation}
    Equations~\eqref{lp} and~\eqref{tp} can be regarded as the parametric equations for $p$ ranging from $1/2$ to $\varepsilon$, which determine the optimal protocol $\tilde{\lambda}(\tilde{t})$. By substituting $\tilde{\lambda}(p)$ into Eq.~\eqref{heat_functional2}, we obtain an expression of the minimal cost
    \begin{equation}\label{Wmin}
          Q_{\text{min}}(\tau,\varepsilon)=-T\Delta S + T\Delta\Sigma_{\text{min}},
    \end{equation}
    where the second term denotes the minimal additional heat production corresponding the minimal entropy production $\Delta\Sigma_{\text{min}}$. Once the jump operators $L_{k}$ are specified, the optimal protocol and minimal thermodynamic cost can be determined by Eqs.~\eqref{lp},~\eqref{tp} and~\eqref{Wmin}, as demonstrated by various examples below. Instead of solving differential equations, which is required for many previous studies~\cite{Ma2022,Dong2024}, our approach involves only solving algebraic equations.

    In the slow-driving regime $\mu\tau\gg1$, the optimal protocol can be explicitly determined by the first-order perturbation~\cite{supp}. The minimal entropy production exhibits the universal $1/\tau$ behavior in this regime. Beyond the slow-driving regime, $\Delta\Sigma_{\text{min}}$ deviates from the $1/\tau$ behavior. Differing from the $1/\tau$ bounds for the additional cost~\cite{Zhen2021, Tan20221}, which are tighter bounds compared to the Landauer principle, Eq.~\eqref{Wmin} represents the tightest bound for the given jump operators. 

    \textit{Two classes.}---\textcolor{black}{We classify thermal environments into two classes based on the convergence or divergence of the rescaled transition rate $R(\tilde{\lambda})$. For the first class, the transition rate from the excited state to the ground state converges to a finite value, which limits the speed of the transition. In contrast, for the second class, the transition rate diverges, which enables arbitrarily fast transitions for sufficiently large driving strength. These two classes exhibit distinct properties for qubit reset.}

    For the first class, we prove~\cite{supp} that the erasure process has a minimal required time for a set error probability or a minimal error probability for a given duration time
    \begin{equation}\label{error}
      \tau>\tau_{\text{min}}=-\frac{\ln{2\varepsilon}}{\mu}~~\text{or}~~\varepsilon>\varepsilon_{\text{min}}=\frac{e^{-\mu\tau}}{2},
    \end{equation}
    which suggests that the arbitrarily small error probability can not be achieved in a finite time for such systems. Physically, the infimum corresponds to the quench protocol $\tilde{\lambda}(\tilde{t})\rightarrow+\infty$, resulting in the fastest decrease of the population of the excited state and a divergent cost. This inherent trade off means a lower bound to the speed of computation in such cases, which hinders Moore's law to continue, even if infinite heat is generated. A specific example with respect to the fermionic bath is shown in quantum dot~\cite{Diana2013}. Here, we clarify the sufficient condition for Eq.~\eqref{error}, which is unrestricted to the specific form of the jump operators for the first class. In contrast, the second class does not have such a constraint.

    \textit{Fast-driving regime.}---Different from the slow-driving regime, the behavior of minimal entropy production for fast qubit reset depends on the class. For the first class, where the fast-driving regime means that the duration time approaches the minimal required time for a set error probability $\tau\rightarrow\tau_{\text{min}}$, we find~\cite{supp} the minimal entropy production exhibits logarithmic behavior
    \begin{equation}\label{fast_bounded}
      \Delta\Sigma_{\min} = O\left( \ln\dfrac{1}{\mu(\tau - \tau_{\min})} \right).
    \end{equation}
    \textcolor{black}{Moreover, the optimal protocol shows the similarity in the short-time limit
    \begin{equation}\label{protocol_bounded}
      \tilde{\lambda}(\tilde{t})+\ln{(\tau-\tau_{\min})}=\Lambda(\tilde{t}),
    \end{equation}
    which suggests that the optimal protocol for two different times are related by a translation transformation $\tilde{\lambda}_{1}(\tilde{t})+\ln{(\tau_{1}-\tau_{\min})}=\tilde{\lambda}_{2}(\tilde{t})+\ln{(\tau_{2}-\tau_{\min})}$ for two different times. Here, $\Lambda(\tilde{t})$ can be considered as a renormalized optimal protocol.} 
    
    For the second class, the minimal required time vanished $\tau_{\min}\rightarrow 0$. The behavior of the minimal entropy production in the fast-driving regime can be calculated by 
    \begin{equation}\label{fast_unbounded}
      \Delta\Sigma_{\text{min}}= O\left(\phi_{0}^{-1}\left[\frac{1}{c_{0}\mu\tau}\left(\frac{1}{2\varepsilon}-1\right)\right]\right),
    \end{equation}
    where $\phi_{0}$ and $c_{0}$ represent the leading-order terms of the asymptotic expansion $R(\tilde{\lambda})=\sum_{n}c_{n}\phi_{n}(\tilde{\lambda})$ as $\tilde{\lambda}\rightarrow +\infty$. \textcolor{black}{The optimal protocol satisfies transformation in the fast-driving regime
    \begin{equation}\label{protocol_unbounded}
      c_{0}\phi_{0}[\lambda(\tilde{t})]\tau=\Lambda(\tilde{t}),
    \end{equation}
    which implies that the optimal protocol for two different time can be related by a transformation $\phi_{0}[\lambda_{1}(\tilde{t})]\tau_{1}=\phi_{0}[\lambda_{2}(\tilde{t})]\tau_{2}$.}
    
    \textcolor{black}{The quantitative estimation of protocols and energy lower limits required in real-world computation is important for planning the design of future computer architectures. Our results suggest that, for fast qubit reset, one can determine the minimal heat dissipation and the optimal protocol for a shorter reset time based on the results for a longer reset time. Different from resetting the classical bits described by the overdamped Langevin dynamics in the double-well potential~\cite{Proesmans2020L, Proesmans2020E, Boyd2022}, the results of qubit depend significantly on the properties of the heat bath. By coarse-grained the continuous double-well model into a discrete two-level system~\cite{Zhen2021} with the specific $R(\tilde{\lambda})$, the results of resetting classical bit can be obtained from our results.}

    \begin{figure}[ht]
      \centering
      \includegraphics[width=1\linewidth]{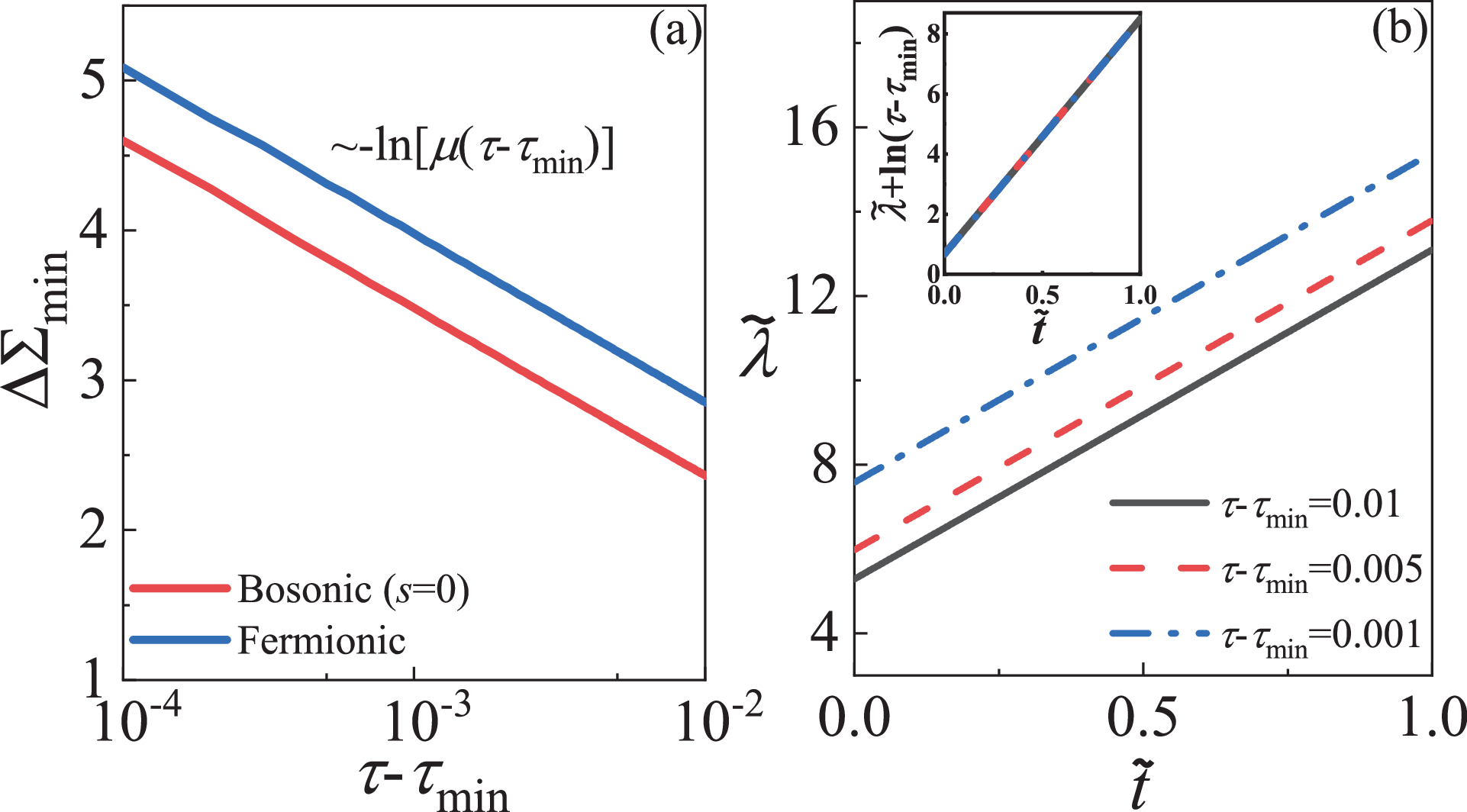}
      \caption{\label{Fig2} \textcolor{black}{(a) The minimal entropy production as the function of duration time with semi-log scale for the bosonic heat bath with $s=0$ (red solid line) and the fermionic heat bath (blue solid line) in the fast-driving regime. The black dash line is drawn for $\ln{(\tau-\tau_{\min})}$ as reference. (b) The optimal protocols $\tilde{\lambda}(\tilde{t})$ for the fermionic heat bath for different times $\tau-\tau_{\min}=0.01$ (black solid line), $\tau-\tau_{\min}=0.005$ (red dash line) and $\tau-\tau_{\min}=0.001$ (blue dash dot line). The inset shows the optimal protocols overlap after transformation $\tilde{\lambda}(\tilde{t})+\ln{(\tau-\tau_{\min})}$. Here, $\varepsilon=0.01$, $\beta=1$ and $\mu=1$.}}
      \end{figure}


    \textit{Examples.}---To demonstrate our results, we consider a qubit coupled to the fermionic and bosonic heat baths, respectively. For the fermionic bath which can be realized by the quantum dots~\cite{Scandi2022}, the jump operators are given by $L_{0}=\sqrt{\mu(1-N_{\text{FD}})}|0\rangle\langle1|$ and $L_{1}=\sqrt{\mu N_{\text{FD}}}|1\rangle\langle0|$, where $N_{\text{FD}}:=1/(e^{\beta\lambda}+1)$ is the Fermi-Dirac distribution. The jump operators of the bosonic heat bath read $L_{0}=\sqrt{\mu\lambda^{s}(1+N_{\text{BE}})}|0\rangle\langle1|$ and $L_{1}=\sqrt{\mu\lambda^{s}N_{\text{BE}}}|1\rangle\langle0|$, where ${\lambda}^{s}$ represents the bath spectral density and $N_{\text{BE}}:=1/(e^{\beta\lambda}-1)$ is the Bose-Einstein distribution. The spectral density exponent $s\in[0,1)$, $s=1$, and $s>1$ correspond to the sub-Ohmic, Ohmic, and super-Ohmic cases, respectively. The fermionic bath and the bosonic bath with $s=0$ belong to the first class, while the bosonic bath with $s\neq0$ belongs to the second class.

    In Fig.~\ref{Fig2}(a), we display the minimal entropy production as the function of time in the fast-driving regime for the bosonic heat bath with $s=0$ and the fermionic heat bath. As predicted by Eq.~\eqref{fast_bounded}, both cases exhibit the logarithmic divergence in the short-time limit. As shown in Fig.~\ref{Fig2}(b), the optimal protocol is discontinuous at the initial moment $\tilde{\lambda}(0^{+})\neq\tilde{\lambda}(0)=0$, which is a generic feature of optimal protocols~\cite{Tim2007, Salazar2019, Proesmans2020L}. Besides, the quench at $t=0$ can be experimentally realized~\cite{Barker2022, Scandi2022}. \textcolor{black}{The inset of Fig.~\ref{Fig2}(b) shows that the optimal protocols for different reset times perfectly overlap after the transformation $\tilde{\lambda}(\tilde{t})+\ln{(\tau-\tau_{\min})}$, consistent with Eq.~\eqref{protocol_bounded}.}

    \begin{figure}[ht]
      \centering
      \includegraphics[width=0.9\linewidth]{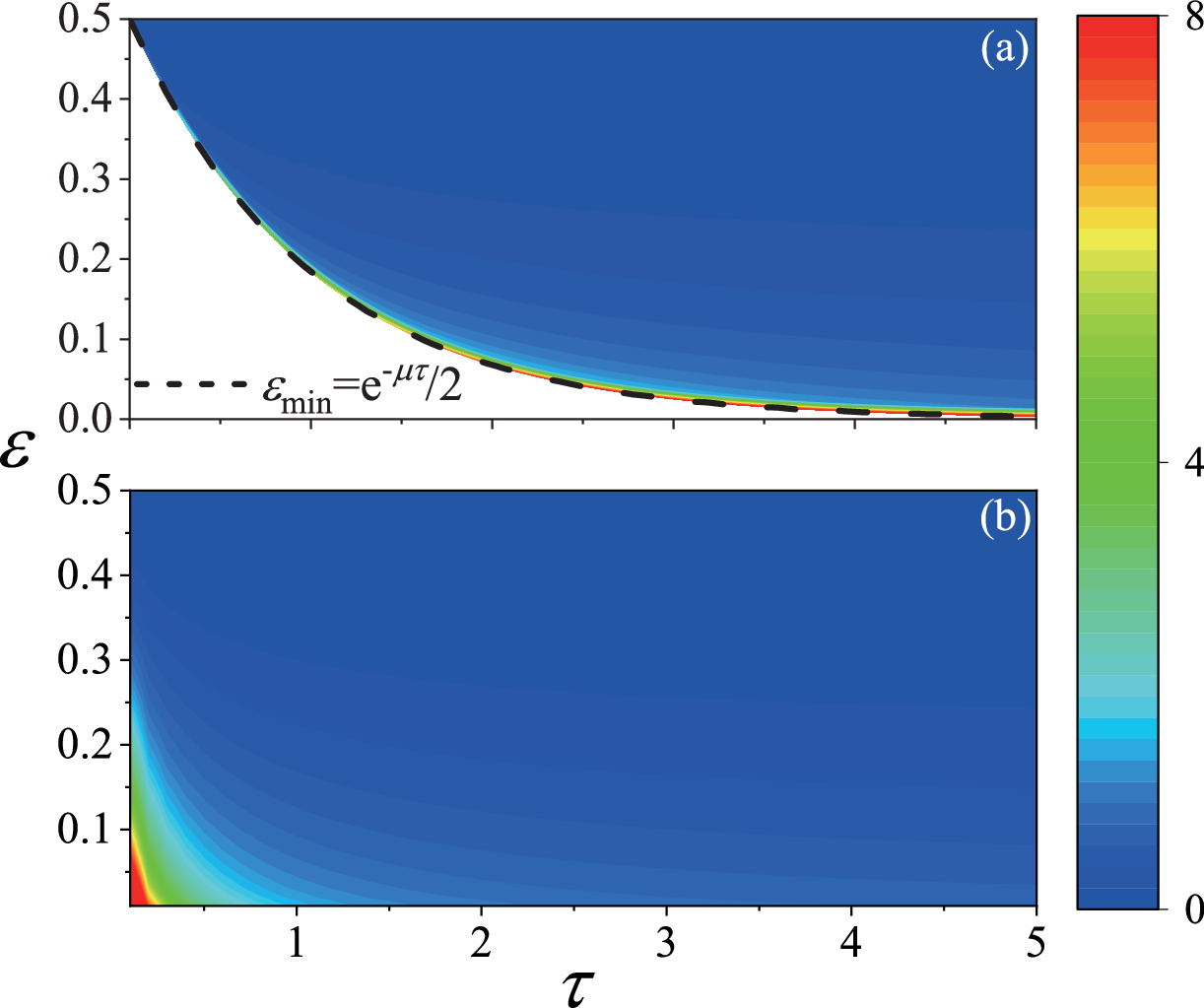}
      \caption{\label{Fig3} Contour plot of the minimal heat production as the function of error probability and erasure time for (a) \textcolor{black}{the quantum dot} and (b) the bosonic heat bath with $s=1$. The black dashed line in (a) represents the minimal error probability for a given erasure time Eq.~\eqref{error}. Here, $\beta=1$ and $\mu=1$.}
      \end{figure}

    We prove~\cite{supp} that the minimal heat production of the bosonic bath with $s=0$ sets the lower bound of the first class. Therefore, the bosonic bath has advantages over the fermionic bath for qubit reset, attributed to the statistical properties of bosons, which facilitate more energy transfer channels at lower energy levels. Conversely, fermions, constrained by the Pauli exclusion principle, exhibit less efficient in energy transfer. \textcolor{black}{This difference can be regarded as a quantum effect distinct from the quantum coherence.} Our findings contrast with those for quantum refrigerators, where fermionic heat baths have been shown to be advantageous over bosonic ones~\cite{Damas2023}.
    
    The different between the two classes is clearly shown in Fig.~\ref{Fig3}. Figure~\ref{Fig3}(a) illustrates the contour plot of the minimal heat production as a function of the duration time $\tau$ and the error probability $\varepsilon$ for the quantum dot belonging to the first class. The boundary perfectly dovetails with Eq.~\eqref{error}. The contour plot for the bosonic bath with $s=1$ belonging to the second class is shown in Fig.~\ref{Fig3}(b), where an arbitrarily small error probability can be achieved in an arbitrarily short time with a sufficient energy cost.
    \begin{figure}[ht]
      \centering
      \includegraphics[width=1\linewidth]{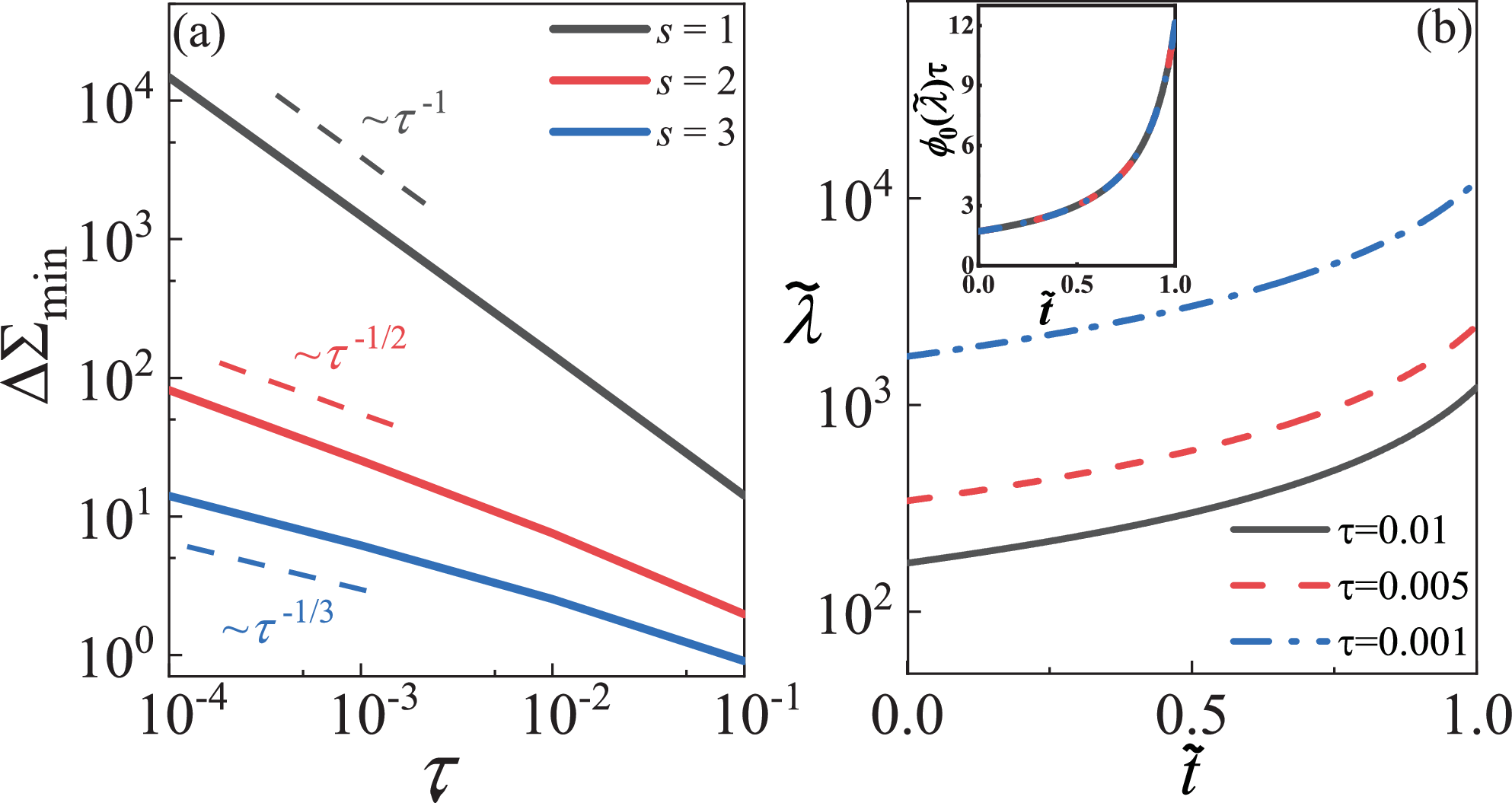}
      \caption{\label{Fig4} \textcolor{black}{(a) The minimal entropy production as a function of duration time with log-log scale for the bosonic bath with $s=1$ (black solid lines), $s=2$ (red solid lines) and $s=3$ (blue solid lines). The three dash lines are drawn for $\tau^{-1}$ (black), $\tau^{-\frac{1}{2}}$ (red) and $\tau^{-\frac{1}{3}}$ (blue) as references. (b) The optimal protocols $\tilde{\lambda}(\tilde{t})$ for the bosonic heat bath with $s=1$ for different times $\tau=0.01$ (black solid line), $\tau=0.005$ (red dash line) and $\tau=0.001$ (blue dash dot line). The inset shows the optimal protocols overlap after transformation $\phi_{0}[\lambda(\tilde{t})]\tau$.}}
      \end{figure}

    Our general formulas extend the study of the bosonic baths with different spectral densities to arbitrary reset time, going beyond the slow-driving regime studied by the geometrical method~\cite{Ma2022}. For the bosonic heat bath with $s\neq 0$, the rescaled transition rate exhibits asymptotic expansion $R(\tilde{\lambda})={\tilde{\lambda}}^{s}+o({\tilde{\lambda}}^{s})$, which corresponds to $\phi_{0}(\tilde{\lambda})={\tilde{\lambda}}^{s}$ and $c_{0}=1$ in Eq.~\eqref{fast_unbounded}. As a result, the behavior of the entropy production in the fast-driving regime reads
    \begin{equation}\label{fast_bosonic}
      \Delta\Sigma_{\min} = O\left[(\mu\tau)^{-\frac{1}{s}}\right],
    \end{equation}
    which suggests a scaling divergence in the additional heat production as the reset time decreases. Figure~\ref{Fig4}(a) illustrates the minimal entropy production for different values of $s$ in the fast-driving regime, confirming the above results. \textcolor{black}{The reset speed $1/\tau$ can be regarded as the ``thermodynamic force'' of the nonequilibrium reset process. One can obtain the linear response behavior for the Ohmic case $s=1$, which is the same as the result of classical bit in the short-time limit. The sub-Ohmic and super-Ohmic cases generate more and less heat dissipation than the Ohmic case, which resemble the capacitive and inductive circuits respectively. Equation~\eqref{fast_bosonic} indicates that, for the super-Ohmic case, the time-energy cost $Q\tau$ that is often considered in applications~\cite{Gaudenzi2018,Buffoni2023} tends to zero in the short-time limit, while for the Ohmic and sub-Ohmic cases, the time-energy cost exists a non-zero lower bound. Our results suggest the quantum advantage for the super-Ohmic case, which results from the quantum properties of the heat bath. Furthermore, according to Eq.~\eqref{protocol_unbounded}, the optimal protocol for the bosonic heat bath with $s\neq 0$ satisfies
    \begin{equation}\label{similarity}
      \tilde{\lambda}^{s}(\tilde{t})\tau=\Lambda(\tilde{t}).
    \end{equation}
    The case of $s=1$ exemplifies this similarity in the Fig.~\ref{Fig4}(b), where the optimal protocols in the fast-driving regime with different reset times overlap after the transformation $\tilde{\lambda}^{s}(\tilde{t})\tau$.}

    \textit{Summary.}---In summary, we develop a general framework for the minimal thermodynamic cost and the optimal protocol for qubit reset with arbitrary reset time. We study the properties of optimal qubit reset in the short-time limit based on this framework. \textcolor{black}{Our findings suggest that in practical devices, thermal environments where the jump operator converges should be avoided as much as possible and the bosonic heat bath with super-Ohmic spectral density are well-suited for qubit reset. The standard CMOS technology will reach its fundamental physical limits in the next two decades~\cite{Ho2023}. Our result can help the continuation of Moore's law after this limit is reached.} Our study provides a framework for other fast-driving quantum information processes, such as quantum gate operations~\cite{Tan2024} and quantum Szilard heat engine~\cite{Zhou2024}.

    \begin{acknowledgments}
    \textit{Acknowledgments.}---This work was financially supported from the National Natural Science Foundation of China (Grants No. 12075199, No. 12247172 and No. 12347151), Natural Science Foundation of Fujian Province (Grant No. 2021J01006) and Jiangxi Province (No. 20212BAB201024).
    \end{acknowledgments}


\begin{thebibliography}{52}%
  \makeatletter
  \providecommand \@ifxundefined [1]{%
   \@ifx{#1\undefined}
  }%
  \providecommand \@ifnum [1]{%
   \ifnum #1\expandafter \@firstoftwo
   \else \expandafter \@secondoftwo
   \fi
  }%
  \providecommand \@ifx [1]{%
   \ifx #1\expandafter \@firstoftwo
   \else \expandafter \@secondoftwo
   \fi
  }%
  \providecommand \natexlab [1]{#1}%
  \providecommand \enquote  [1]{``#1''}%
  \providecommand \bibnamefont  [1]{#1}%
  \providecommand \bibfnamefont [1]{#1}%
  \providecommand \citenamefont [1]{#1}%
  \providecommand \href@noop [0]{\@secondoftwo}%
  \providecommand \href [0]{\begingroup \@sanitize@url \@href}%
  \providecommand \@href[1]{\@@startlink{#1}\@@href}%
  \providecommand \@@href[1]{\endgroup#1\@@endlink}%
  \providecommand \@sanitize@url [0]{\catcode `\\12\catcode `\$12\catcode
    `\&12\catcode `\#12\catcode `\^12\catcode `\_12\catcode `\%12\relax}%
  \providecommand \@@startlink[1]{}%
  \providecommand \@@endlink[0]{}%
  \providecommand \url  [0]{\begingroup\@sanitize@url \@url }%
  \providecommand \@url [1]{\endgroup\@href {#1}{\urlprefix }}%
  \providecommand \urlprefix  [0]{URL }%
  \providecommand \Eprint [0]{\href }%
  \providecommand \doibase [0]{http://dx.doi.org/}%
  \providecommand \selectlanguage [0]{\@gobble}%
  \providecommand \bibinfo  [0]{\@secondoftwo}%
  \providecommand \bibfield  [0]{\@secondoftwo}%
  \providecommand \translation [1]{[#1]}%
  \providecommand \BibitemOpen [0]{}%
  \providecommand \bibitemStop [0]{}%
  \providecommand \bibitemNoStop [0]{.\EOS\space}%
  \providecommand \EOS [0]{\spacefactor3000\relax}%
  \providecommand \BibitemShut  [1]{\csname bibitem#1\endcsname}%
  \let\auto@bib@innerbib\@empty
  \bibitem [{\citenamefont {DiVincenzo}(2000)}]{Di2000}%
    \BibitemOpen
    \bibfield  {author} {\bibinfo {author} {\bibfnamefont {D.~P.}\ \bibnamefont
    {DiVincenzo}},\ }\href {\doibase
    https://doi.org/10.1002/1521-3978(200009)48:9/11<771::AID-PROP771>3.0.CO;2-E}
    {\bibfield  {journal} {\bibinfo  {journal} {Fortschr. Phys.}\ }\textbf
    {\bibinfo {volume} {48}},\ \bibinfo {pages} {771} (\bibinfo {year}
    {2000})}\BibitemShut {NoStop}%
  \bibitem [{\citenamefont {Landauer}(1961)}]{Landauer1961}%
    \BibitemOpen
    \bibfield  {author} {\bibinfo {author} {\bibfnamefont {R.}~\bibnamefont
    {Landauer}},\ }\href {\doibase 10.1147/rd.53.0183} {\bibfield  {journal}
    {\bibinfo  {journal} {IBM J. Res. Dev.}\ }\textbf {\bibinfo {volume} {5}},\
    \bibinfo {pages} {183} (\bibinfo {year} {1961})}\BibitemShut {NoStop}%
  \bibitem [{\citenamefont {Bennett}(1982)}]{Bennett1982}%
    \BibitemOpen
    \bibfield  {author} {\bibinfo {author} {\bibfnamefont {C.~H.}\ \bibnamefont
    {Bennett}},\ }\href {https://link.springer.com/article/10.1007/BF02084158}
    {\bibfield  {journal} {\bibinfo  {journal} {Int. J. Theor. Phys.}\ }\textbf
    {\bibinfo {volume} {21}},\ \bibinfo {pages} {905} (\bibinfo {year}
    {1982})}\BibitemShut {NoStop}%
  \bibitem [{\citenamefont {Georgescu}(2021)}]{Georgescu2021}%
    \BibitemOpen
    \bibfield  {author} {\bibinfo {author} {\bibfnamefont {I.}~\bibnamefont
    {Georgescu}},\ }\href {\doibase 10.1038/s42254-021-00400-8} {\bibfield
    {journal} {\bibinfo  {journal} {Nat. Rev. Phys.}\ }\textbf {\bibinfo {volume}
    {3}},\ \bibinfo {pages} {770} (\bibinfo {year} {2021})}\BibitemShut {NoStop}%
  \bibitem [{\citenamefont {Parrondo}\ \emph {et~al.}(2015)\citenamefont
    {Parrondo}, \citenamefont {Horowitz},\ and\ \citenamefont
    {Sagawa}}]{Parrondo2015}%
    \BibitemOpen
    \bibfield  {author} {\bibinfo {author} {\bibfnamefont {J.~M.~R.}\
    \bibnamefont {Parrondo}}, \bibinfo {author} {\bibfnamefont {J.~M.}\
    \bibnamefont {Horowitz}}, \ and\ \bibinfo {author} {\bibfnamefont
    {T.}~\bibnamefont {Sagawa}},\ }\href {\doibase 10.1038/nphys3230} {\bibfield
    {journal} {\bibinfo  {journal} {Nat. Phys.}\ }\textbf {\bibinfo {volume}
    {11}},\ \bibinfo {pages} {131} (\bibinfo {year} {2015})}\BibitemShut
    {NoStop}%
  \bibitem [{\citenamefont {B{\'e}rut}\ \emph {et~al.}(2012)\citenamefont
    {B{\'e}rut}, \citenamefont {Arakelyan}, \citenamefont {Petrosyan},
    \citenamefont {Ciliberto}, \citenamefont {Dillenschneider},\ and\
    \citenamefont {Lutz}}]{Antoine2012}%
    \BibitemOpen
    \bibfield  {author} {\bibinfo {author} {\bibfnamefont {A.}~\bibnamefont
    {B{\'e}rut}}, \bibinfo {author} {\bibfnamefont {A.}~\bibnamefont
    {Arakelyan}}, \bibinfo {author} {\bibfnamefont {A.}~\bibnamefont
    {Petrosyan}}, \bibinfo {author} {\bibfnamefont {S.}~\bibnamefont
    {Ciliberto}}, \bibinfo {author} {\bibfnamefont {R.}~\bibnamefont
    {Dillenschneider}}, \ and\ \bibinfo {author} {\bibfnamefont {E.}~\bibnamefont
    {Lutz}},\ }\href {\doibase 10.1038/nature10872} {\bibfield  {journal}
    {\bibinfo  {journal} {Nature}\ }\textbf {\bibinfo {volume} {483}},\ \bibinfo
    {pages} {187} (\bibinfo {year} {2012})}\BibitemShut {NoStop}%
  \bibitem [{\citenamefont {Jun}\ \emph {et~al.}(2014)\citenamefont {Jun},
    \citenamefont {Gavrilov},\ and\ \citenamefont {Bechhoefer}}]{Jun2014}%
    \BibitemOpen
    \bibfield  {author} {\bibinfo {author} {\bibfnamefont {Y.}~\bibnamefont
    {Jun}}, \bibinfo {author} {\bibfnamefont {M.}~\bibnamefont {Gavrilov}}, \
    and\ \bibinfo {author} {\bibfnamefont {J.}~\bibnamefont {Bechhoefer}},\
    }\href {\doibase 10.1103/PhysRevLett.113.190601} {\bibfield  {journal}
    {\bibinfo  {journal} {Phys. Rev. Lett.}\ }\textbf {\bibinfo {volume} {113}},\
    \bibinfo {pages} {190601} (\bibinfo {year} {2014})}\BibitemShut {NoStop}%
  \bibitem [{\citenamefont {Gavrilov}\ and\ \citenamefont
    {Bechhoefer}(2016)}]{Gavrilov2016}%
    \BibitemOpen
    \bibfield  {author} {\bibinfo {author} {\bibfnamefont {M.}~\bibnamefont
    {Gavrilov}}\ and\ \bibinfo {author} {\bibfnamefont {J.}~\bibnamefont
    {Bechhoefer}},\ }\href {\doibase 10.1103/PhysRevLett.117.200601} {\bibfield
    {journal} {\bibinfo  {journal} {Phys. Rev. Lett.}\ }\textbf {\bibinfo
    {volume} {117}},\ \bibinfo {pages} {200601} (\bibinfo {year}
    {2016})}\BibitemShut {NoStop}%
  \bibitem [{\citenamefont {Yan}\ \emph {et~al.}(2018)\citenamefont {Yan},
    \citenamefont {Xiong}, \citenamefont {Rehan}, \citenamefont {Zhou},
    \citenamefont {Liang}, \citenamefont {Chen}, \citenamefont {Zhang},
    \citenamefont {Yang}, \citenamefont {Ma},\ and\ \citenamefont
    {Feng}}]{Yan2018}%
    \BibitemOpen
    \bibfield  {author} {\bibinfo {author} {\bibfnamefont {L.~L.}\ \bibnamefont
    {Yan}}, \bibinfo {author} {\bibfnamefont {T.~P.}\ \bibnamefont {Xiong}},
    \bibinfo {author} {\bibfnamefont {K.}~\bibnamefont {Rehan}}, \bibinfo
    {author} {\bibfnamefont {F.}~\bibnamefont {Zhou}}, \bibinfo {author}
    {\bibfnamefont {D.~F.}\ \bibnamefont {Liang}}, \bibinfo {author}
    {\bibfnamefont {L.}~\bibnamefont {Chen}}, \bibinfo {author} {\bibfnamefont
    {J.~Q.}\ \bibnamefont {Zhang}}, \bibinfo {author} {\bibfnamefont {W.~L.}\
    \bibnamefont {Yang}}, \bibinfo {author} {\bibfnamefont {Z.~H.}\ \bibnamefont
    {Ma}}, \ and\ \bibinfo {author} {\bibfnamefont {M.}~\bibnamefont {Feng}},\
    }\href {\doibase 10.1103/PhysRevLett.120.210601} {\bibfield  {journal}
    {\bibinfo  {journal} {Phys. Rev. Lett.}\ }\textbf {\bibinfo {volume} {120}},\
    \bibinfo {pages} {210601} (\bibinfo {year} {2018})}\BibitemShut {NoStop}%
  \bibitem [{\citenamefont {Hong}\ \emph {et~al.}(2016)\citenamefont {Hong},
    \citenamefont {Lambson}, \citenamefont {Dhuey},\ and\ \citenamefont
    {Bokor}}]{Hong2016}%
    \BibitemOpen
    \bibfield  {author} {\bibinfo {author} {\bibfnamefont {J.}~\bibnamefont
    {Hong}}, \bibinfo {author} {\bibfnamefont {B.}~\bibnamefont {Lambson}},
    \bibinfo {author} {\bibfnamefont {S.}~\bibnamefont {Dhuey}}, \ and\ \bibinfo
    {author} {\bibfnamefont {J.}~\bibnamefont {Bokor}},\ }\href {\doibase
    10.1126/sciadv.1501492} {\bibfield  {journal} {\bibinfo  {journal} {Sci.
    Adv.}\ }\textbf {\bibinfo {volume} {2}},\ \bibinfo {pages} {e1501492}
    (\bibinfo {year} {2016})}\BibitemShut {NoStop}%
  \bibitem [{\citenamefont {Martini}\ \emph {et~al.}(2016)\citenamefont
    {Martini}, \citenamefont {Pancaldi}, \citenamefont {Madami}, \citenamefont
    {Vavassori}, \citenamefont {Gubbiotti}, \citenamefont {Tacchi}, \citenamefont
    {Hartmann}, \citenamefont {Emmerling}, \citenamefont {Höfling},
    \citenamefont {Worschech},\ and\ \citenamefont {Carlotti}}]{Martini2016}%
    \BibitemOpen
    \bibfield  {author} {\bibinfo {author} {\bibfnamefont {L.}~\bibnamefont
    {Martini}}, \bibinfo {author} {\bibfnamefont {M.}~\bibnamefont {Pancaldi}},
    \bibinfo {author} {\bibfnamefont {M.}~\bibnamefont {Madami}}, \bibinfo
    {author} {\bibfnamefont {P.}~\bibnamefont {Vavassori}}, \bibinfo {author}
    {\bibfnamefont {G.}~\bibnamefont {Gubbiotti}}, \bibinfo {author}
    {\bibfnamefont {S.}~\bibnamefont {Tacchi}}, \bibinfo {author} {\bibfnamefont
    {F.}~\bibnamefont {Hartmann}}, \bibinfo {author} {\bibfnamefont
    {M.}~\bibnamefont {Emmerling}}, \bibinfo {author} {\bibfnamefont
    {S.}~\bibnamefont {Höfling}}, \bibinfo {author} {\bibfnamefont
    {L.}~\bibnamefont {Worschech}}, \ and\ \bibinfo {author} {\bibfnamefont
    {G.}~\bibnamefont {Carlotti}},\ }\href {\doibase
    https://doi.org/10.1016/j.nanoen.2015.10.028} {\bibfield  {journal} {\bibinfo
     {journal} {Nano Energy}\ }\textbf {\bibinfo {volume} {19}},\ \bibinfo
    {pages} {108} (\bibinfo {year} {2016})}\BibitemShut {NoStop}%
  \bibitem [{\citenamefont {Gaudenzi}\ \emph {et~al.}(2018)\citenamefont
    {Gaudenzi}, \citenamefont {Burzur{\'i}}, \citenamefont {Maegawa},
    \citenamefont {van~der Zant},\ and\ \citenamefont {Luis}}]{Gaudenzi2018}%
    \BibitemOpen
    \bibfield  {author} {\bibinfo {author} {\bibfnamefont {R.}~\bibnamefont
    {Gaudenzi}}, \bibinfo {author} {\bibfnamefont {E.}~\bibnamefont
    {Burzur{\'i}}}, \bibinfo {author} {\bibfnamefont {S.}~\bibnamefont
    {Maegawa}}, \bibinfo {author} {\bibfnamefont {H.~S.~J.}\ \bibnamefont
    {van~der Zant}}, \ and\ \bibinfo {author} {\bibfnamefont {F.}~\bibnamefont
    {Luis}},\ }\href {\doibase 10.1038/s41567-018-0070-7} {\bibfield  {journal}
    {\bibinfo  {journal} {Nat. Phys.}\ }\textbf {\bibinfo {volume} {14}},\
    \bibinfo {pages} {565} (\bibinfo {year} {2018})}\BibitemShut {NoStop}%
  \bibitem [{\citenamefont {Dago}\ \emph {et~al.}(2021)\citenamefont {Dago},
    \citenamefont {Pereda}, \citenamefont {Barros}, \citenamefont {Ciliberto},\
    and\ \citenamefont {Bellon}}]{Dago2021}%
    \BibitemOpen
    \bibfield  {author} {\bibinfo {author} {\bibfnamefont {S.}~\bibnamefont
    {Dago}}, \bibinfo {author} {\bibfnamefont {J.}~\bibnamefont {Pereda}},
    \bibinfo {author} {\bibfnamefont {N.}~\bibnamefont {Barros}}, \bibinfo
    {author} {\bibfnamefont {S.}~\bibnamefont {Ciliberto}}, \ and\ \bibinfo
    {author} {\bibfnamefont {L.}~\bibnamefont {Bellon}},\ }\href {\doibase
    10.1103/PhysRevLett.126.170601} {\bibfield  {journal} {\bibinfo  {journal}
    {Phys. Rev. Lett.}\ }\textbf {\bibinfo {volume} {126}},\ \bibinfo {pages}
    {170601} (\bibinfo {year} {2021})}\BibitemShut {NoStop}%
  \bibitem [{\citenamefont {Dago}\ and\ \citenamefont {Bellon}(2022)}]{Dago2022}%
    \BibitemOpen
    \bibfield  {author} {\bibinfo {author} {\bibfnamefont {S.}~\bibnamefont
    {Dago}}\ and\ \bibinfo {author} {\bibfnamefont {L.}~\bibnamefont {Bellon}},\
    }\href {\doibase 10.1103/PhysRevLett.128.070604} {\bibfield  {journal}
    {\bibinfo  {journal} {Phys. Rev. Lett.}\ }\textbf {\bibinfo {volume} {128}},\
    \bibinfo {pages} {070604} (\bibinfo {year} {2022})}\BibitemShut {NoStop}%
  \bibitem [{\citenamefont {Dago}\ \emph {et~al.}(2023)\citenamefont {Dago},
    \citenamefont {Ciliberto},\ and\ \citenamefont {Bellon}}]{Dago2023}%
    \BibitemOpen
    \bibfield  {author} {\bibinfo {author} {\bibfnamefont {S.}~\bibnamefont
    {Dago}}, \bibinfo {author} {\bibfnamefont {S.}~\bibnamefont {Ciliberto}}, \
    and\ \bibinfo {author} {\bibfnamefont {L.}~\bibnamefont {Bellon}},\ }\href
    {\doibase 10.1073/pnas.2301742120} {\bibfield  {journal} {\bibinfo  {journal}
    {Proc. Natl. Acad. Sci. U.S.A.}\ }\textbf {\bibinfo {volume} {120}},\
    \bibinfo {pages} {e2301742120} (\bibinfo {year} {2023})}\BibitemShut
    {NoStop}%
  \bibitem [{\citenamefont {Scandi}\ \emph {et~al.}(2022)\citenamefont {Scandi},
    \citenamefont {Barker}, \citenamefont {Lehmann}, \citenamefont {Dick},
    \citenamefont {Maisi},\ and\ \citenamefont {Perarnau-Llobet}}]{Scandi2022}%
    \BibitemOpen
    \bibfield  {author} {\bibinfo {author} {\bibfnamefont {M.}~\bibnamefont
    {Scandi}}, \bibinfo {author} {\bibfnamefont {D.}~\bibnamefont {Barker}},
    \bibinfo {author} {\bibfnamefont {S.}~\bibnamefont {Lehmann}}, \bibinfo
    {author} {\bibfnamefont {K.~A.}\ \bibnamefont {Dick}}, \bibinfo {author}
    {\bibfnamefont {V.~F.}\ \bibnamefont {Maisi}}, \ and\ \bibinfo {author}
    {\bibfnamefont {M.}~\bibnamefont {Perarnau-Llobet}},\ }\href {\doibase
    10.1103/PhysRevLett.129.270601} {\bibfield  {journal} {\bibinfo  {journal}
    {Phys. Rev. Lett.}\ }\textbf {\bibinfo {volume} {129}},\ \bibinfo {pages}
    {270601} (\bibinfo {year} {2022})}\BibitemShut {NoStop}%
  \textcolor{black}{\bibitem [{\citenamefont {Chu}\ and\ \citenamefont {Cai}(2022)}]{Chu2022}%
    \BibitemOpen
    \bibfield  {author} {\bibinfo {author} {\bibfnamefont {Y.}~\bibnamefont
    {Chu}}\ and\ \bibinfo {author} {\bibfnamefont {J.}~\bibnamefont {Cai}},\
    }\href {\doibase 10.1103/PhysRevLett.128.200501} {\bibfield  {journal}
    {\bibinfo  {journal} {Phys. Rev. Lett.}\ }\textbf {\bibinfo {volume} {128}},\
    \bibinfo {pages} {200501} (\bibinfo {year} {2022})}\BibitemShut {NoStop}%
  \bibitem [{\citenamefont {Niroula}\ \emph {et~al.}(2024)\citenamefont
    {Niroula}, \citenamefont {Dolde}, \citenamefont {Zheng}, \citenamefont
    {Bringewatt}, \citenamefont {Ehrenberg}, \citenamefont {Cox}, \citenamefont
    {Thompson}, \citenamefont {Gullans}, \citenamefont {Kolkowitz},\ and\
    \citenamefont {Gorshkov}}]{Niroula2024}%
    \BibitemOpen
    \bibfield  {author} {\bibinfo {author} {\bibfnamefont {P.}~\bibnamefont
    {Niroula}}, \bibinfo {author} {\bibfnamefont {J.}~\bibnamefont {Dolde}},
    \bibinfo {author} {\bibfnamefont {X.}~\bibnamefont {Zheng}}, \bibinfo
    {author} {\bibfnamefont {J.}~\bibnamefont {Bringewatt}}, \bibinfo {author}
    {\bibfnamefont {A.}~\bibnamefont {Ehrenberg}}, \bibinfo {author}
    {\bibfnamefont {K.~C.}\ \bibnamefont {Cox}}, \bibinfo {author} {\bibfnamefont
    {J.}~\bibnamefont {Thompson}}, \bibinfo {author} {\bibfnamefont {M.~J.}\
    \bibnamefont {Gullans}}, \bibinfo {author} {\bibfnamefont {S.}~\bibnamefont
    {Kolkowitz}}, \ and\ \bibinfo {author} {\bibfnamefont {A.~V.}\ \bibnamefont
    {Gorshkov}},\ }\href {\doibase 10.1103/PhysRevLett.133.080801} {\bibfield
    {journal} {\bibinfo  {journal} {Phys. Rev. Lett.}\ }\textbf {\bibinfo
    {volume} {133}},\ \bibinfo {pages} {080801} (\bibinfo {year}
    {2024})}\BibitemShut {NoStop}%
  \bibitem [{\citenamefont {Kempes}\ \emph {et~al.}(2017)\citenamefont {Kempes},
    \citenamefont {Wolpert}, \citenamefont {Cohen},\ and\ \citenamefont
    {P{\'e}rez-Mercader}}]{kempes2017}%
    \BibitemOpen
    \bibfield  {author} {\bibinfo {author} {\bibfnamefont {C.~P.}\ \bibnamefont
    {Kempes}}, \bibinfo {author} {\bibfnamefont {D.}~\bibnamefont {Wolpert}},
    \bibinfo {author} {\bibfnamefont {Z.}~\bibnamefont {Cohen}}, \ and\ \bibinfo
    {author} {\bibfnamefont {J.}~\bibnamefont {P{\'e}rez-Mercader}},\ }\href@noop
    {} {\bibfield  {journal} {\bibinfo  {journal} {Phil. Trans. R. Soc. A.}\
    }\textbf {\bibinfo {volume} {375}},\ \bibinfo {pages} {20160343} (\bibinfo
    {year} {2017})}\BibitemShut {NoStop}%
  \bibitem [{\citenamefont {Wolpert}\ \emph {et~al.}(2024)\citenamefont
    {Wolpert}, \citenamefont {Korbel}, \citenamefont {Lynn}, \citenamefont
    {Tasnim}, \citenamefont {Grochow}, \citenamefont {Karde{\c{s}}},
    \citenamefont {Aimone}, \citenamefont {Balasubramanian}, \citenamefont
    {De~Giuli}, \citenamefont {Doty}, \citenamefont {Freitas}, \citenamefont
    {Marsili}, \citenamefont {Ouldridge}, \citenamefont {Richa}, \citenamefont
    {Riechers}, \citenamefont {Rold{\'a}n}, \citenamefont {Rubenstein},
    \citenamefont {Toroczkai},\ and\ \citenamefont {Paradiso}}]{Wolpert2024}%
    \BibitemOpen
    \bibfield  {author} {\bibinfo {author} {\bibfnamefont {D.~H.}\ \bibnamefont
    {Wolpert}}, \bibinfo {author} {\bibfnamefont {J.}~\bibnamefont {Korbel}},
    \bibinfo {author} {\bibfnamefont {C.~W.}\ \bibnamefont {Lynn}}, \bibinfo
    {author} {\bibfnamefont {F.}~\bibnamefont {Tasnim}}, \bibinfo {author}
    {\bibfnamefont {J.~A.}\ \bibnamefont {Grochow}}, \bibinfo {author}
    {\bibfnamefont {G.}~\bibnamefont {Karde{\c{s}}}}, \bibinfo {author}
    {\bibfnamefont {J.~B.}\ \bibnamefont {Aimone}}, \bibinfo {author}
    {\bibfnamefont {V.}~\bibnamefont {Balasubramanian}}, \bibinfo {author}
    {\bibfnamefont {E.}~\bibnamefont {De~Giuli}}, \bibinfo {author}
    {\bibfnamefont {D.}~\bibnamefont {Doty}}, \bibinfo {author} {\bibfnamefont
    {N.}~\bibnamefont {Freitas}}, \bibinfo {author} {\bibfnamefont
    {M.}~\bibnamefont {Marsili}}, \bibinfo {author} {\bibfnamefont {T.~E.}\
    \bibnamefont {Ouldridge}}, \bibinfo {author} {\bibfnamefont {A.~W.}\
    \bibnamefont {Richa}}, \bibinfo {author} {\bibfnamefont {P.}~\bibnamefont
    {Riechers}}, \bibinfo {author} {\bibfnamefont {{\'E}.}~\bibnamefont
    {Rold{\'a}n}}, \bibinfo {author} {\bibfnamefont {B.}~\bibnamefont
    {Rubenstein}}, \bibinfo {author} {\bibfnamefont {Z.}~\bibnamefont
    {Toroczkai}}, \ and\ \bibinfo {author} {\bibfnamefont {J.}~\bibnamefont
    {Paradiso}},\ }\href {\doibase 10.1073/pnas.2321112121} {\bibfield  {journal}
    {\bibinfo  {journal} {Proc. Natl. Acad. Sci. U.S.A.}\ }\textbf {\bibinfo
    {volume} {121}},\ \bibinfo {pages} {e2321112121} (\bibinfo {year}
    {2024})}\BibitemShut {NoStop}%
  \bibitem [{\citenamefont {Waldrop}(2016)}]{Waldrop2016}%
    \BibitemOpen
    \bibfield  {author} {\bibinfo {author} {\bibfnamefont {M.~M.}\ \bibnamefont
    {Waldrop}},\ }\href {\doibase 10.1038/530144a} {\bibfield  {journal}
    {\bibinfo  {journal} {Nature}\ }\textbf {\bibinfo {volume} {530}},\ \bibinfo
    {pages} {144} (\bibinfo {year} {2016})}\BibitemShut {NoStop}}%
  \bibitem [{\citenamefont {Zhen}\ \emph {et~al.}(2021)\citenamefont {Zhen},
    \citenamefont {Egloff}, \citenamefont {Modi},\ and\ \citenamefont
    {Dahlsten}}]{Zhen2021}%
    \BibitemOpen
    \bibfield  {author} {\bibinfo {author} {\bibfnamefont {Y.-Z.}\ \bibnamefont
    {Zhen}}, \bibinfo {author} {\bibfnamefont {D.}~\bibnamefont {Egloff}},
    \bibinfo {author} {\bibfnamefont {K.}~\bibnamefont {Modi}}, \ and\ \bibinfo
    {author} {\bibfnamefont {O.}~\bibnamefont {Dahlsten}},\ }\href {\doibase
    10.1103/PhysRevLett.127.190602} {\bibfield  {journal} {\bibinfo  {journal}
    {Phys. Rev. Lett.}\ }\textbf {\bibinfo {volume} {127}},\ \bibinfo {pages}
    {190602} (\bibinfo {year} {2021})}\BibitemShut {NoStop}%
  \bibitem [{\citenamefont {Van~Vu}\ and\ \citenamefont
    {Saito}(2022)}]{Tan20221}%
    \BibitemOpen
    \bibfield  {author} {\bibinfo {author} {\bibfnamefont {T.}~\bibnamefont
    {Van~Vu}}\ and\ \bibinfo {author} {\bibfnamefont {K.}~\bibnamefont {Saito}},\
    }\href {\doibase 10.1103/PhysRevLett.128.010602} {\bibfield  {journal}
    {\bibinfo  {journal} {Phys. Rev. Lett.}\ }\textbf {\bibinfo {volume} {128}},\
    \bibinfo {pages} {010602} (\bibinfo {year} {2022})}\BibitemShut {NoStop}%
  \bibitem [{\citenamefont {Zhen}\ \emph {et~al.}(2022)\citenamefont {Zhen},
    \citenamefont {Egloff}, \citenamefont {Modi},\ and\ \citenamefont
    {Dahlsten}}]{Zhen2022}%
    \BibitemOpen
    \bibfield  {author} {\bibinfo {author} {\bibfnamefont {Y.-Z.}\ \bibnamefont
    {Zhen}}, \bibinfo {author} {\bibfnamefont {D.}~\bibnamefont {Egloff}},
    \bibinfo {author} {\bibfnamefont {K.}~\bibnamefont {Modi}}, \ and\ \bibinfo
    {author} {\bibfnamefont {O.}~\bibnamefont {Dahlsten}},\ }\href {\doibase
    10.1103/PhysRevE.105.044147} {\bibfield  {journal} {\bibinfo  {journal}
    {Phys. Rev. E}\ }\textbf {\bibinfo {volume} {105}},\ \bibinfo {pages}
    {044147} (\bibinfo {year} {2022})}\BibitemShut {NoStop}%
  \bibitem [{\citenamefont {Lee}\ \emph {et~al.}(2022)\citenamefont {Lee},
    \citenamefont {Lee}, \citenamefont {Kwon},\ and\ \citenamefont
    {Park}}]{Lee2022}%
    \BibitemOpen
    \bibfield  {author} {\bibinfo {author} {\bibfnamefont {J.~S.}\ \bibnamefont
    {Lee}}, \bibinfo {author} {\bibfnamefont {S.}~\bibnamefont {Lee}}, \bibinfo
    {author} {\bibfnamefont {H.}~\bibnamefont {Kwon}}, \ and\ \bibinfo {author}
    {\bibfnamefont {H.}~\bibnamefont {Park}},\ }\href {\doibase
    10.1103/PhysRevLett.129.120603} {\bibfield  {journal} {\bibinfo  {journal}
    {Phys. Rev. Lett.}\ }\textbf {\bibinfo {volume} {129}},\ \bibinfo {pages}
    {120603} (\bibinfo {year} {2022})}\BibitemShut {NoStop}%
  \bibitem [{\citenamefont {Van~Vu}\ and\ \citenamefont {Saito}(2023)}]{Tan2023}%
    \BibitemOpen
    \bibfield  {author} {\bibinfo {author} {\bibfnamefont {T.}~\bibnamefont
    {Van~Vu}}\ and\ \bibinfo {author} {\bibfnamefont {K.}~\bibnamefont {Saito}},\
    }\href {\doibase 10.1103/PhysRevX.13.011013} {\bibfield  {journal} {\bibinfo
    {journal} {Phys. Rev. X}\ }\textbf {\bibinfo {volume} {13}},\ \bibinfo
    {pages} {011013} (\bibinfo {year} {2023})}\BibitemShut {NoStop}%
  \bibitem [{\citenamefont {Proesmans}\ \emph
    {et~al.}(2020{\natexlab{a}})\citenamefont {Proesmans}, \citenamefont
    {Ehrich},\ and\ \citenamefont {Bechhoefer}}]{Proesmans2020L}%
    \BibitemOpen
    \bibfield  {author} {\bibinfo {author} {\bibfnamefont {K.}~\bibnamefont
    {Proesmans}}, \bibinfo {author} {\bibfnamefont {J.}~\bibnamefont {Ehrich}}, \
    and\ \bibinfo {author} {\bibfnamefont {J.}~\bibnamefont {Bechhoefer}},\
    }\href {\doibase 10.1103/PhysRevLett.125.100602} {\bibfield  {journal}
    {\bibinfo  {journal} {Phys. Rev. Lett.}\ }\textbf {\bibinfo {volume} {125}},\
    \bibinfo {pages} {100602} (\bibinfo {year} {2020}{\natexlab{a}})}\BibitemShut
    {NoStop}%
  \bibitem [{\citenamefont {Proesmans}\ \emph
    {et~al.}(2020{\natexlab{b}})\citenamefont {Proesmans}, \citenamefont
    {Ehrich},\ and\ \citenamefont {Bechhoefer}}]{Proesmans2020E}%
    \BibitemOpen
    \bibfield  {author} {\bibinfo {author} {\bibfnamefont {K.}~\bibnamefont
    {Proesmans}}, \bibinfo {author} {\bibfnamefont {J.}~\bibnamefont {Ehrich}}, \
    and\ \bibinfo {author} {\bibfnamefont {J.}~\bibnamefont {Bechhoefer}},\
    }\href {\doibase 10.1103/PhysRevE.102.032105} {\bibfield  {journal} {\bibinfo
     {journal} {Phys. Rev. E}\ }\textbf {\bibinfo {volume} {102}},\ \bibinfo
    {pages} {032105} (\bibinfo {year} {2020}{\natexlab{b}})}\BibitemShut
    {NoStop}%
  \bibitem [{\citenamefont {Boyd}\ \emph {et~al.}(2022)\citenamefont {Boyd},
    \citenamefont {Patra}, \citenamefont {Jarzynski},\ and\ \citenamefont
    {Crutchfield}}]{Boyd2022}%
    \BibitemOpen
    \bibfield  {author} {\bibinfo {author} {\bibfnamefont {A.~B.}\ \bibnamefont
    {Boyd}}, \bibinfo {author} {\bibfnamefont {A.}~\bibnamefont {Patra}},
    \bibinfo {author} {\bibfnamefont {C.}~\bibnamefont {Jarzynski}}, \ and\
    \bibinfo {author} {\bibfnamefont {J.~P.}\ \bibnamefont {Crutchfield}},\
    }\href {\doibase 10.1007/s10955-022-02871-0} {\bibfield  {journal} {\bibinfo
    {journal} {J. Stat. Phys.}\ }\textbf {\bibinfo {volume} {187}},\ \bibinfo
    {pages} {17} (\bibinfo {year} {2022})}\BibitemShut {NoStop}%
  \bibitem [{\citenamefont {Konopik}\ \emph {et~al.}(2023)\citenamefont
    {Konopik}, \citenamefont {Korten}, \citenamefont {Lutz},\ and\ \citenamefont
    {Linke}}]{Konopik2023}%
    \BibitemOpen
    \bibfield  {author} {\bibinfo {author} {\bibfnamefont {M.}~\bibnamefont
    {Konopik}}, \bibinfo {author} {\bibfnamefont {T.}~\bibnamefont {Korten}},
    \bibinfo {author} {\bibfnamefont {E.}~\bibnamefont {Lutz}}, \ and\ \bibinfo
    {author} {\bibfnamefont {H.}~\bibnamefont {Linke}},\ }\href {\doibase
    10.1038/s41467-023-36020-2} {\bibfield  {journal} {\bibinfo  {journal}
    {Nature Communications}\ }\textbf {\bibinfo {volume} {14}},\ \bibinfo {pages}
    {447} (\bibinfo {year} {2023})}\BibitemShut {NoStop}%
  \bibitem [{\citenamefont {Ma}\ \emph {et~al.}(2020)\citenamefont {Ma},
    \citenamefont {Zhai}, \citenamefont {Chen}, \citenamefont {Sun},\ and\
    \citenamefont {Dong}}]{Ma2020}%
    \BibitemOpen
    \bibfield  {author} {\bibinfo {author} {\bibfnamefont {Y.-H.}\ \bibnamefont
    {Ma}}, \bibinfo {author} {\bibfnamefont {R.-X.}\ \bibnamefont {Zhai}},
    \bibinfo {author} {\bibfnamefont {J.}~\bibnamefont {Chen}}, \bibinfo {author}
    {\bibfnamefont {C.~P.}\ \bibnamefont {Sun}}, \ and\ \bibinfo {author}
    {\bibfnamefont {H.}~\bibnamefont {Dong}},\ }\href {\doibase
    10.1103/PhysRevLett.125.210601} {\bibfield  {journal} {\bibinfo  {journal}
    {Phys. Rev. Lett.}\ }\textbf {\bibinfo {volume} {125}},\ \bibinfo {pages}
    {210601} (\bibinfo {year} {2020})}\BibitemShut {NoStop}%
  \bibitem [{\citenamefont {Yuan}\ \emph {et~al.}(2022)\citenamefont {Yuan},
    \citenamefont {Ma},\ and\ \citenamefont {Sun}}]{Yuan2022}%
    \BibitemOpen
    \bibfield  {author} {\bibinfo {author} {\bibfnamefont {H.}~\bibnamefont
    {Yuan}}, \bibinfo {author} {\bibfnamefont {Y.-H.}\ \bibnamefont {Ma}}, \ and\
    \bibinfo {author} {\bibfnamefont {C.~P.}\ \bibnamefont {Sun}},\ }\href
    {\doibase 10.1103/PhysRevE.105.L022101} {\bibfield  {journal} {\bibinfo
    {journal} {Phys. Rev. E}\ }\textbf {\bibinfo {volume} {105}},\ \bibinfo
    {pages} {L022101} (\bibinfo {year} {2022})}\BibitemShut {NoStop}%
  \bibitem [{\citenamefont {Ma}\ \emph {et~al.}(2022)\citenamefont {Ma},
    \citenamefont {Chen}, \citenamefont {Sun},\ and\ \citenamefont
    {Dong}}]{Ma2022}%
    \BibitemOpen
    \bibfield  {author} {\bibinfo {author} {\bibfnamefont {Y.-H.}\ \bibnamefont
    {Ma}}, \bibinfo {author} {\bibfnamefont {J.-F.}\ \bibnamefont {Chen}},
    \bibinfo {author} {\bibfnamefont {C.~P.}\ \bibnamefont {Sun}}, \ and\
    \bibinfo {author} {\bibfnamefont {H.}~\bibnamefont {Dong}},\ }\href {\doibase
    10.1103/PhysRevE.106.034112} {\bibfield  {journal} {\bibinfo  {journal}
    {Phys. Rev. E}\ }\textbf {\bibinfo {volume} {106}},\ \bibinfo {pages}
    {034112} (\bibinfo {year} {2022})}\BibitemShut {NoStop}%
  \textcolor{black}{\bibitem [{\citenamefont {Magnard}\ \emph {et~al.}(2018)\citenamefont
    {Magnard}, \citenamefont {Kurpiers}, \citenamefont {Royer}, \citenamefont
    {Walter}, \citenamefont {Besse}, \citenamefont {Gasparinetti}, \citenamefont
    {Pechal}, \citenamefont {Heinsoo}, \citenamefont {Storz}, \citenamefont
    {Blais},\ and\ \citenamefont {Wallraff}}]{Magnard2018}%
    \BibitemOpen
    \bibfield  {author} {\bibinfo {author} {\bibfnamefont {P.}~\bibnamefont
    {Magnard}}, \bibinfo {author} {\bibfnamefont {P.}~\bibnamefont {Kurpiers}},
    \bibinfo {author} {\bibfnamefont {B.}~\bibnamefont {Royer}}, \bibinfo
    {author} {\bibfnamefont {T.}~\bibnamefont {Walter}}, \bibinfo {author}
    {\bibfnamefont {J.-C.}\ \bibnamefont {Besse}}, \bibinfo {author}
    {\bibfnamefont {S.}~\bibnamefont {Gasparinetti}}, \bibinfo {author}
    {\bibfnamefont {M.}~\bibnamefont {Pechal}}, \bibinfo {author} {\bibfnamefont
    {J.}~\bibnamefont {Heinsoo}}, \bibinfo {author} {\bibfnamefont
    {S.}~\bibnamefont {Storz}}, \bibinfo {author} {\bibfnamefont
    {A.}~\bibnamefont {Blais}}, \ and\ \bibinfo {author} {\bibfnamefont
    {A.}~\bibnamefont {Wallraff}},\ }\href {\doibase
    10.1103/PhysRevLett.121.060502} {\bibfield  {journal} {\bibinfo  {journal}
    {Phys. Rev. Lett.}\ }\textbf {\bibinfo {volume} {121}},\ \bibinfo {pages}
    {060502} (\bibinfo {year} {2018})}\BibitemShut {NoStop}%
  \bibitem [{\citenamefont {Zhou}\ \emph {et~al.}(2021)\citenamefont {Zhou},
    \citenamefont {Zhang}, \citenamefont {Yin}, \citenamefont {Huai},
    \citenamefont {Gu}, \citenamefont {Xu}, \citenamefont {Allcock},
    \citenamefont {Liu}, \citenamefont {Xi}, \citenamefont {Yu}, \citenamefont
    {Zhang}, \citenamefont {Zhang}, \citenamefont {Li}, \citenamefont {Song},
    \citenamefont {Wang}, \citenamefont {Zheng}, \citenamefont {An},
    \citenamefont {Zheng},\ and\ \citenamefont {Zhang}}]{Zhou2021}%
    \BibitemOpen
    \bibfield  {author} {\bibinfo {author} {\bibfnamefont {Y.}~\bibnamefont
    {Zhou}}, \bibinfo {author} {\bibfnamefont {Z.}~\bibnamefont {Zhang}},
    \bibinfo {author} {\bibfnamefont {Z.}~\bibnamefont {Yin}}, \bibinfo {author}
    {\bibfnamefont {S.}~\bibnamefont {Huai}}, \bibinfo {author} {\bibfnamefont
    {X.}~\bibnamefont {Gu}}, \bibinfo {author} {\bibfnamefont {X.}~\bibnamefont
    {Xu}}, \bibinfo {author} {\bibfnamefont {J.}~\bibnamefont {Allcock}},
    \bibinfo {author} {\bibfnamefont {F.}~\bibnamefont {Liu}}, \bibinfo {author}
    {\bibfnamefont {G.}~\bibnamefont {Xi}}, \bibinfo {author} {\bibfnamefont
    {Q.}~\bibnamefont {Yu}}, \bibinfo {author} {\bibfnamefont {H.}~\bibnamefont
    {Zhang}}, \bibinfo {author} {\bibfnamefont {M.}~\bibnamefont {Zhang}},
    \bibinfo {author} {\bibfnamefont {H.}~\bibnamefont {Li}}, \bibinfo {author}
    {\bibfnamefont {X.}~\bibnamefont {Song}}, \bibinfo {author} {\bibfnamefont
    {Z.}~\bibnamefont {Wang}}, \bibinfo {author} {\bibfnamefont {D.}~\bibnamefont
    {Zheng}}, \bibinfo {author} {\bibfnamefont {S.}~\bibnamefont {An}}, \bibinfo
    {author} {\bibfnamefont {Y.}~\bibnamefont {Zheng}}, \ and\ \bibinfo {author}
    {\bibfnamefont {S.}~\bibnamefont {Zhang}},\ }\href {\doibase
    10.1038/s41467-021-26205-y} {\bibfield  {journal} {\bibinfo  {journal} {Nat.
    Commun.}\ }\textbf {\bibinfo {volume} {12}},\ \bibinfo {pages} {5924}
    (\bibinfo {year} {2021})}\BibitemShut {NoStop}%
  \bibitem [{\citenamefont {Johnson}\ \emph {et~al.}(2022)\citenamefont
    {Johnson}, \citenamefont {M\k{a}dzik}, \citenamefont {Hudson}, \citenamefont
    {Itoh}, \citenamefont {Jakob}, \citenamefont {Jamieson}, \citenamefont
    {Dzurak},\ and\ \citenamefont {Morello}}]{Johnson2022}%
    \BibitemOpen
    \bibfield  {author} {\bibinfo {author} {\bibfnamefont {M.~A.~I.}\
    \bibnamefont {Johnson}}, \bibinfo {author} {\bibfnamefont {M.~T.}\
    \bibnamefont {M\k{a}dzik}}, \bibinfo {author} {\bibfnamefont {F.~E.}\
    \bibnamefont {Hudson}}, \bibinfo {author} {\bibfnamefont {K.~M.}\
    \bibnamefont {Itoh}}, \bibinfo {author} {\bibfnamefont {A.~M.}\ \bibnamefont
    {Jakob}}, \bibinfo {author} {\bibfnamefont {D.~N.}\ \bibnamefont {Jamieson}},
    \bibinfo {author} {\bibfnamefont {A.}~\bibnamefont {Dzurak}}, \ and\ \bibinfo
    {author} {\bibfnamefont {A.}~\bibnamefont {Morello}},\ }\href {\doibase
    10.1103/PhysRevX.12.041008} {\bibfield  {journal} {\bibinfo  {journal} {Phys.
    Rev. X}\ }\textbf {\bibinfo {volume} {12}},\ \bibinfo {pages} {041008}
    (\bibinfo {year} {2022})}\BibitemShut {NoStop}%
  \bibitem [{\citenamefont {Reiner}\ \emph {et~al.}(2024)\citenamefont {Reiner},
    \citenamefont {Chung}, \citenamefont {Misha}, \citenamefont {Lehner},
    \citenamefont {Moehle}, \citenamefont {Poulos}, \citenamefont {Monir},
    \citenamefont {Charde}, \citenamefont {Macha}, \citenamefont {Kranz},
    \citenamefont {Thorvaldson}, \citenamefont {Thorgrimsson}, \citenamefont
    {Keith}, \citenamefont {Hsueh}, \citenamefont {Rahman}, \citenamefont
    {Gorman}, \citenamefont {Keizer},\ and\ \citenamefont
    {Simmons}}]{Reiner2024}%
    \BibitemOpen
    \bibfield  {author} {\bibinfo {author} {\bibfnamefont {J.}~\bibnamefont
    {Reiner}}, \bibinfo {author} {\bibfnamefont {Y.}~\bibnamefont {Chung}},
    \bibinfo {author} {\bibfnamefont {S.~H.}\ \bibnamefont {Misha}}, \bibinfo
    {author} {\bibfnamefont {C.}~\bibnamefont {Lehner}}, \bibinfo {author}
    {\bibfnamefont {C.}~\bibnamefont {Moehle}}, \bibinfo {author} {\bibfnamefont
    {D.}~\bibnamefont {Poulos}}, \bibinfo {author} {\bibfnamefont
    {S.}~\bibnamefont {Monir}}, \bibinfo {author} {\bibfnamefont {K.~J.}\
    \bibnamefont {Charde}}, \bibinfo {author} {\bibfnamefont {P.}~\bibnamefont
    {Macha}}, \bibinfo {author} {\bibfnamefont {L.}~\bibnamefont {Kranz}},
    \bibinfo {author} {\bibfnamefont {I.}~\bibnamefont {Thorvaldson}}, \bibinfo
    {author} {\bibfnamefont {B.}~\bibnamefont {Thorgrimsson}}, \bibinfo {author}
    {\bibfnamefont {D.}~\bibnamefont {Keith}}, \bibinfo {author} {\bibfnamefont
    {Y.~L.}\ \bibnamefont {Hsueh}}, \bibinfo {author} {\bibfnamefont
    {R.}~\bibnamefont {Rahman}}, \bibinfo {author} {\bibfnamefont {S.~K.}\
    \bibnamefont {Gorman}}, \bibinfo {author} {\bibfnamefont {J.~G.}\
    \bibnamefont {Keizer}}, \ and\ \bibinfo {author} {\bibfnamefont {M.~Y.}\
    \bibnamefont {Simmons}},\ }\href {\doibase 10.1038/s41565-023-01596-9}
    {\bibfield  {journal} {\bibinfo  {journal} {Nat. Nanotechnol.}\ }\textbf
    {\bibinfo {volume} {19}},\ \bibinfo {pages} {605} (\bibinfo {year}
    {2024})}\BibitemShut {NoStop}%
  \bibitem [{\citenamefont {Wang}\ \emph {et~al.}(2024)\citenamefont {Wang},
    \citenamefont {Wu}, \citenamefont {Wang}, \citenamefont {Ma}, \citenamefont
    {Zhang}, \citenamefont {Chen}, \citenamefont {Deng}, \citenamefont {Gao},
    \citenamefont {Hu}, \citenamefont {Ma}, \citenamefont {Song}, \citenamefont
    {Xia}, \citenamefont {Ying}, \citenamefont {Zhan}, \citenamefont {Zhao},\
    and\ \citenamefont {Deng}}]{Alibaba2024}%
    \BibitemOpen
    \bibfield  {author} {\bibinfo {author} {\bibfnamefont {T.}~\bibnamefont
    {Wang}}, \bibinfo {author} {\bibfnamefont {F.}~\bibnamefont {Wu}}, \bibinfo
    {author} {\bibfnamefont {F.}~\bibnamefont {Wang}}, \bibinfo {author}
    {\bibfnamefont {X.}~\bibnamefont {Ma}}, \bibinfo {author} {\bibfnamefont
    {G.}~\bibnamefont {Zhang}}, \bibinfo {author} {\bibfnamefont
    {J.}~\bibnamefont {Chen}}, \bibinfo {author} {\bibfnamefont {H.}~\bibnamefont
    {Deng}}, \bibinfo {author} {\bibfnamefont {R.}~\bibnamefont {Gao}}, \bibinfo
    {author} {\bibfnamefont {R.}~\bibnamefont {Hu}}, \bibinfo {author}
    {\bibfnamefont {L.}~\bibnamefont {Ma}}, \bibinfo {author} {\bibfnamefont
    {Z.}~\bibnamefont {Song}}, \bibinfo {author} {\bibfnamefont {T.}~\bibnamefont
    {Xia}}, \bibinfo {author} {\bibfnamefont {M.}~\bibnamefont {Ying}}, \bibinfo
    {author} {\bibfnamefont {H.}~\bibnamefont {Zhan}}, \bibinfo {author}
    {\bibfnamefont {H.-H.}\ \bibnamefont {Zhao}}, \ and\ \bibinfo {author}
    {\bibfnamefont {C.}~\bibnamefont {Deng}},\ }\href {\doibase
    10.1103/PhysRevLett.132.230601} {\bibfield  {journal} {\bibinfo  {journal}
    {Phys. Rev. Lett.}\ }\textbf {\bibinfo {volume} {132}},\ \bibinfo {pages}
    {230601} (\bibinfo {year} {2024})}\BibitemShut {NoStop}}%
  \bibitem [{\citenamefont {Browne}\ \emph {et~al.}(2014)\citenamefont {Browne},
    \citenamefont {Garner}, \citenamefont {Dahlsten},\ and\ \citenamefont
    {Vedral}}]{Browne2014}%
    \BibitemOpen
    \bibfield  {author} {\bibinfo {author} {\bibfnamefont {C.}~\bibnamefont
    {Browne}}, \bibinfo {author} {\bibfnamefont {A.~J.~P.}\ \bibnamefont
    {Garner}}, \bibinfo {author} {\bibfnamefont {O.~C.~O.}\ \bibnamefont
    {Dahlsten}}, \ and\ \bibinfo {author} {\bibfnamefont {V.}~\bibnamefont
    {Vedral}},\ }\href {\doibase 10.1103/PhysRevLett.113.100603} {\bibfield
    {journal} {\bibinfo  {journal} {Phys. Rev. Lett.}\ }\textbf {\bibinfo
    {volume} {113}},\ \bibinfo {pages} {100603} (\bibinfo {year}
    {2014})}\BibitemShut {NoStop}%
  \bibitem [{\citenamefont {Miller}\ \emph {et~al.}(2020)\citenamefont {Miller},
    \citenamefont {Guarnieri}, \citenamefont {Mitchison},\ and\ \citenamefont
    {Goold}}]{Miller2020}%
    \BibitemOpen
    \bibfield  {author} {\bibinfo {author} {\bibfnamefont {H.~J.~D.}\
    \bibnamefont {Miller}}, \bibinfo {author} {\bibfnamefont {G.}~\bibnamefont
    {Guarnieri}}, \bibinfo {author} {\bibfnamefont {M.~T.}\ \bibnamefont
    {Mitchison}}, \ and\ \bibinfo {author} {\bibfnamefont {J.}~\bibnamefont
    {Goold}},\ }\href {\doibase 10.1103/PhysRevLett.125.160602} {\bibfield
    {journal} {\bibinfo  {journal} {Phys. Rev. Lett.}\ }\textbf {\bibinfo
    {volume} {125}},\ \bibinfo {pages} {160602} (\bibinfo {year}
    {2020})}\BibitemShut {NoStop}%
  \textcolor{black}{\bibitem [{\citenamefont {Plastina}\ \emph {et~al.}(2014)\citenamefont
    {Plastina}, \citenamefont {Alecce}, \citenamefont {Apollaro}, \citenamefont
    {Falcone}, \citenamefont {Francica}, \citenamefont {Galve}, \citenamefont
    {Lo~Gullo},\ and\ \citenamefont {Zambrini}}]{Plastina2014}%
    \BibitemOpen
    \bibfield  {author} {\bibinfo {author} {\bibfnamefont {F.}~\bibnamefont
    {Plastina}}, \bibinfo {author} {\bibfnamefont {A.}~\bibnamefont {Alecce}},
    \bibinfo {author} {\bibfnamefont {T.~J.~G.}\ \bibnamefont {Apollaro}},
    \bibinfo {author} {\bibfnamefont {G.}~\bibnamefont {Falcone}}, \bibinfo
    {author} {\bibfnamefont {G.}~\bibnamefont {Francica}}, \bibinfo {author}
    {\bibfnamefont {F.}~\bibnamefont {Galve}}, \bibinfo {author} {\bibfnamefont
    {N.}~\bibnamefont {Lo~Gullo}}, \ and\ \bibinfo {author} {\bibfnamefont
    {R.}~\bibnamefont {Zambrini}},\ }\href {\doibase
    10.1103/PhysRevLett.113.260601} {\bibfield  {journal} {\bibinfo  {journal}
    {Phys. Rev. Lett.}\ }\textbf {\bibinfo {volume} {113}},\ \bibinfo {pages}
    {260601} (\bibinfo {year} {2014})}\BibitemShut {NoStop}}%
  \bibitem [{sup()}]{supp}%
    \BibitemOpen
    \href@noop {} {\ }\bibinfo {note} {See the Supplementary
    Material}\BibitemShut {NoStop}%
  \bibitem [{\citenamefont {Huang}\ \emph {et~al.}(2024)\citenamefont {Huang},
    \citenamefont {Li},\ and\ \citenamefont {Dong}}]{Dong2024}%
    \BibitemOpen
    \bibfield  {author} {\bibinfo {author} {\bibfnamefont {H.-B.}\ \bibnamefont
    {Huang}}, \bibinfo {author} {\bibfnamefont {G.}~\bibnamefont {Li}}, \ and\
    \bibinfo {author} {\bibfnamefont {H.}~\bibnamefont {Dong}},\ }\href {\doibase
    10.1103/PhysRevE.109.064132} {\bibfield  {journal} {\bibinfo  {journal}
    {Phys. Rev. E}\ }\textbf {\bibinfo {volume} {109}},\ \bibinfo {pages}
    {064132} (\bibinfo {year} {2024})}\BibitemShut {NoStop}%
  \bibitem [{\citenamefont {Diana}\ \emph {et~al.}(2013)\citenamefont {Diana},
    \citenamefont {Bagci},\ and\ \citenamefont {Esposito}}]{Diana2013}%
    \BibitemOpen
    \bibfield  {author} {\bibinfo {author} {\bibfnamefont {G.}~\bibnamefont
    {Diana}}, \bibinfo {author} {\bibfnamefont {G.~B.}\ \bibnamefont {Bagci}}, \
    and\ \bibinfo {author} {\bibfnamefont {M.}~\bibnamefont {Esposito}},\ }\href
    {\doibase 10.1103/PhysRevE.87.012111} {\bibfield  {journal} {\bibinfo
    {journal} {Phys. Rev. E}\ }\textbf {\bibinfo {volume} {87}},\ \bibinfo
    {pages} {012111} (\bibinfo {year} {2013})}\BibitemShut {NoStop}%
  \bibitem [{\citenamefont {Schmiedl}\ and\ \citenamefont
    {Seifert}(2007)}]{Tim2007}%
    \BibitemOpen
    \bibfield  {author} {\bibinfo {author} {\bibfnamefont {T.}~\bibnamefont
    {Schmiedl}}\ and\ \bibinfo {author} {\bibfnamefont {U.}~\bibnamefont
    {Seifert}},\ }\href {\doibase 10.1103/PhysRevLett.98.108301} {\bibfield
    {journal} {\bibinfo  {journal} {Phys. Rev. Lett.}\ }\textbf {\bibinfo
    {volume} {98}},\ \bibinfo {pages} {108301} (\bibinfo {year}
    {2007})}\BibitemShut {NoStop}%
  \bibitem [{\citenamefont {Salazar}\ and\ \citenamefont
    {Lira}(2019)}]{Salazar2019}%
    \BibitemOpen
    \bibfield  {author} {\bibinfo {author} {\bibfnamefont {D.~S.~P.}\
    \bibnamefont {Salazar}}\ and\ \bibinfo {author} {\bibfnamefont {S.~A.}\
    \bibnamefont {Lira}},\ }\href {\doibase 10.1103/PhysRevE.99.062119}
    {\bibfield  {journal} {\bibinfo  {journal} {Phys. Rev. E}\ }\textbf {\bibinfo
    {volume} {99}},\ \bibinfo {pages} {062119} (\bibinfo {year}
    {2019})}\BibitemShut {NoStop}%
  \bibitem [{\citenamefont {Barker}\ \emph {et~al.}(2022)\citenamefont {Barker},
    \citenamefont {Scandi}, \citenamefont {Lehmann}, \citenamefont {Thelander},
    \citenamefont {Dick}, \citenamefont {Perarnau-Llobet},\ and\ \citenamefont
    {Maisi}}]{Barker2022}%
    \BibitemOpen
    \bibfield  {author} {\bibinfo {author} {\bibfnamefont {D.}~\bibnamefont
    {Barker}}, \bibinfo {author} {\bibfnamefont {M.}~\bibnamefont {Scandi}},
    \bibinfo {author} {\bibfnamefont {S.}~\bibnamefont {Lehmann}}, \bibinfo
    {author} {\bibfnamefont {C.}~\bibnamefont {Thelander}}, \bibinfo {author}
    {\bibfnamefont {K.~A.}\ \bibnamefont {Dick}}, \bibinfo {author}
    {\bibfnamefont {M.}~\bibnamefont {Perarnau-Llobet}}, \ and\ \bibinfo {author}
    {\bibfnamefont {V.~F.}\ \bibnamefont {Maisi}},\ }\href {\doibase
    10.1103/PhysRevLett.128.040602} {\bibfield  {journal} {\bibinfo  {journal}
    {Phys. Rev. Lett.}\ }\textbf {\bibinfo {volume} {128}},\ \bibinfo {pages}
    {040602} (\bibinfo {year} {2022})}\BibitemShut {NoStop}%
  \bibitem [{\citenamefont {Damas}\ \emph {et~al.}(2023)\citenamefont {Damas},
    \citenamefont {de~Assis},\ and\ \citenamefont {de~Almeida}}]{Damas2023}%
    \BibitemOpen
    \bibfield  {author} {\bibinfo {author} {\bibfnamefont {G.~G.}\ \bibnamefont
    {Damas}}, \bibinfo {author} {\bibfnamefont {R.~J.}\ \bibnamefont {de~Assis}},
    \ and\ \bibinfo {author} {\bibfnamefont {N.~G.}\ \bibnamefont {de~Almeida}},\
    }\href {\doibase 10.1103/PhysRevE.107.034128} {\bibfield  {journal} {\bibinfo
     {journal} {Phys. Rev. E}\ }\textbf {\bibinfo {volume} {107}},\ \bibinfo
    {pages} {034128} (\bibinfo {year} {2023})}\BibitemShut {NoStop}%
  \bibitem [{\citenamefont {Buffoni}\ and\ \citenamefont
    {Campisi}(2023)}]{Buffoni2023}%
    \BibitemOpen
    \bibfield  {author} {\bibinfo {author} {\bibfnamefont {L.}~\bibnamefont
    {Buffoni}}\ and\ \bibinfo {author} {\bibfnamefont {M.}~\bibnamefont
    {Campisi}},\ }\href {\doibase 10.22331/q-2023-03-23-961} {\bibfield
    {journal} {\bibinfo  {journal} {{Quantum}}\ }\textbf {\bibinfo {volume}
    {7}},\ \bibinfo {pages} {961} (\bibinfo {year} {2023})}\BibitemShut {NoStop}%
  \bibitem [{\citenamefont {Ho}\ \emph {et~al.}(2023)\citenamefont {Ho},
    \citenamefont {Erdil},\ and\ \citenamefont {Besiroglu}}]{Ho2023}%
    \BibitemOpen
    \bibfield  {author} {\bibinfo {author} {\bibfnamefont {A.}~\bibnamefont
    {Ho}}, \bibinfo {author} {\bibfnamefont {E.}~\bibnamefont {Erdil}}, \ and\
    \bibinfo {author} {\bibfnamefont {T.}~\bibnamefont {Besiroglu}},\ }in\ \href
    {\doibase 10.1109/ICRC60800.2023.10386559} {\emph {\bibinfo {booktitle} {2023
    IEEE International Conference on Rebooting Computing (ICRC)}}}\ (\bibinfo
    {year} {2023})\BibitemShut {NoStop}%
  \bibitem [{\citenamefont {Van~Vu}\ \emph {et~al.}(2024)\citenamefont {Van~Vu},
    \citenamefont {Kuwahara},\ and\ \citenamefont {Saito}}]{Tan2024}%
    \BibitemOpen
    \bibfield  {author} {\bibinfo {author} {\bibfnamefont {T.}~\bibnamefont
    {Van~Vu}}, \bibinfo {author} {\bibfnamefont {T.}~\bibnamefont {Kuwahara}}, \
    and\ \bibinfo {author} {\bibfnamefont {K.}~\bibnamefont {Saito}},\ }\href
    {\doibase 10.1103/PhysRevResearch.6.033225} {\bibfield  {journal} {\bibinfo
    {journal} {Phys. Rev. Res.}\ }\textbf {\bibinfo {volume} {6}},\ \bibinfo
    {pages} {033225} (\bibinfo {year} {2024})}\BibitemShut {NoStop}%
  \bibitem [{\citenamefont {Zhou}\ \emph {et~al.}(2024)\citenamefont {Zhou},
    \citenamefont {Ma},\ and\ \citenamefont {Sun}}]{Zhou2024}%
    \BibitemOpen
    \bibfield  {author} {\bibinfo {author} {\bibfnamefont {T.-J.}\ \bibnamefont
    {Zhou}}, \bibinfo {author} {\bibfnamefont {Y.-H.}\ \bibnamefont {Ma}}, \ and\
    \bibinfo {author} {\bibfnamefont {C.~P.}\ \bibnamefont {Sun}},\ }\href
    {\doibase 10.1103/PhysRevResearch.6.043001} {\bibfield  {journal} {\bibinfo
    {journal} {Phys. Rev. Res.}\ }\textbf {\bibinfo {volume} {6}},\ \bibinfo
    {pages} {043001} (\bibinfo {year} {2024})}\BibitemShut {NoStop}%
  \end{thebibliography}
%
  
\end{document}